\documentstyle[prl,floats,epsf,aps]{revtex}

\begin{document}

\twocolumn[\hsize\textwidth\columnwidth\hsize
           \csname @twocolumnfalse\endcsname

\title{
Oscillatory Curie temperature of two-dimensional ferromagnets
 }

\author{
M.~Pajda$^{a}$, J.~Kudrnovsk\'y$^{a,b}$, I. Turek$^{c}$, V.~Drchal$^{b}$,
and P.~Bruno$^{a}$
\\
$^a$ Max-Planck Institut f\"ur Mikrostrukturphysik, Weinberg 2,
     D-06120 Halle, Germany
\\
$^b$ Institute of Physics, Academy of Sciences of the Czech Republic,
      \\ Na Slovance 2, CZ-18221 Prague 8, Czech Republic
\\
$^c$ Institute of Physics of Materials, Academy of Sciences of the
       Czech Republic, \\ \v{Z}i\v{z}kova 22, CZ-61662 Brno,
       Czech Republic
\\
 }
\maketitle

\begin{abstract}

The effective exchange interactions of magnetic overlayers
Fe/Cu(001) and Co/Cu(001) covered by a Cu-cap layer of varying
thickness were calculated in real space from first principles.
The effective two-dimensional Heisenberg Hamiltonian was constructed
and used to estimate magnon dispersion laws, spin-wave stiffness
constants, and overlayer Curie temperatures within the mean-field and
random-phase approximations.
Overlayer Curie temperature oscillates as a function of the cap-layer
thickness in a qualitative agreement with a recent experiment.

\end{abstract}

\date{today}

\draft

\pacs{Published in Phys. Rev. Lett. {\bf 85}, 5424 (2000)}

\vskip2pc]

The Curie temperature is one of the most important characteristics of
ferromagnets.
In particular, the Curie temperature of low-dimensional systems such as
ultrathin films is of considerable interest.
In a recent experimental study, Vollmer {\em et al.\/} \cite{Kirsch}
have shown that (i) the Curie temperature of fcc(001)-Fe ultrathin films
on a Cu(001) substrate is considerably modified upon coverage by a Cu-cap
layer, and (ii) that it varies in a non-monotonous manner as a function of
the Cu cap layer thickness, which indicates an oscillatory variation.
An oscillatory behavior of the Curie temperature as a function of the
spacer thickness was also found for fcc(001)-Co/Cu/Ni trilayers \cite{Ney}.
Such a behavior clearly cannot be explained within a localized picture
of magnetism and calls for a first-principles theory of the Curie
temperature in itinerant ferromagnets.
In spite of considerable efforts in last decades a first-principles
calculation of the Curie temperature in the framework of itinerant
magnetism, in particular for low-dimensional systems, remains a very
serious challenge.

One therefore has to rely upon some approximation schemes in order to
calculate the Curie temperature of itinerant ferromagnets.
A particularly simple and yet accurate approach consists in a mapping of
the complicated itinerant electron system onto an effective Heisenberg
model (EHM),
$H = - \sum_{i \neq j} J_{ij}\ {\bf e}_i \cdot {\bf e}_j$,
where ${\bf e}_i$ and ${\bf e}_j$ are the unit vectors of the magnetic
moments at sites $i$ and $j$, and the effective exchange interactions
(EEIs) $J_{ij}$ between any pair of magnetic moments are determined
from first-principles \cite{Lie,Hal,vSA,AS,Szu,Sonia,Marek}.
The thermodynamic properties of the ferromagnet including determination
of the Curie temperature can be then calculated from the EHM by using
statistical mechanical methods.
A simple mean field approximation (MFA) fails in many cases due
to its neglect of
collective excitations (spin-waves), and more sophisticated approximations,
such as the Green function method within the random phase approximation
(GF-RPA) \cite{RPA}, are preferable.
The success of this two-step approach relies upon the fact that it
provides an almost exact description of low-lying magnetic excitations
(spin-waves) which give the largest contribution to the Curie temperature.
On the other hand this approach completely disregards longitudinal
fluctuations of magnetic moments such as the Stoner excitations and it
therefore is not suitable to describe ferromagnets with small exchange
splitting such as, e.g., fcc-Ni, in which exist Stoner excitations with
a rather low energy.
We have recently applied this approach to bulk bcc-Fe, fcc-Co, and fcc-Ni
and obtained a reasonable agreement with experimental Curie temperatures
of Fe and Co, but not for Ni \cite{Marek} similarly as in recent
calculations based on the adiabatic spin-wave theory \cite{Hal} or an
alternative first-principles theory of spin fluctuations based on idea
of generalized Onsager cavity field \cite{DLM}.

In the present paper we wish to calculate exchange interaction parameters,
spin-wave stiffness constants, and Curie temperatures of two-dimensional
monolayers of Fe and Co and, in particular, to investigate the influence
of the substrate and of the cap layer.
We find that (i) the exchange parameters, spin-wave stiffness constants,
and Curie temperatures are strongly modified by the presence of a
metallic substrate and/or a cap layer, and that (ii) they exhibit an
oscillatory variation with the cap layer thickness.
This behavior is due to the Ruderman-Kittel-Kasuya-Yoshida (RKKY) character
of exchange interactions in itinerant ferromagnets.
Our results are in a good qualitative agreement with the observations of
Vollmer {\em et al.\/} \cite{Kirsch} for which they provide the most
natural explanation.
The same theory can be used to interprete the experiment of Ney {\em et al.\/}
\cite{Ney}, but the detailed analysis deserves a separate study.

The electronic structure of the system was determined in the framework of
the first principles tight-binding linear muffin-tin orbital method (TB-LMTO)
generalized to surfaces \cite{book}.
In the framework of the magnetic force-theorem \cite{Lie,Dede}, the
expression for the EEIs between two sites $i$ and $j$ anywhere in the
system is \cite{Lie,Antr}
\begin{equation}
J_{ij} =\frac{1}{4\pi}\int_{C} \, {\rm Im \, tr}_L \,
\left\{ \delta_{i}(z) \, g^{\uparrow}_{ij}(z) \, \delta_{j}(z) \,
g^{\downarrow}_{ji}(z) \right\} \, {\rm d} z \, .
\label{eq_Jij}
\end{equation}
Here ${\rm tr}_L$ denotes the trace over the angular momentum $L=(\ell m)$,
$\delta_{i}(z)=P_{i}^{\uparrow}(z)-P_{i}^{\downarrow}(z)$ where
$P^{\sigma}_i(z)$ are $L$-diagonal matrices of potential functions
of the TB-LMTO method ($\sigma=\uparrow, \downarrow$), energy integration
is performed in the upper half of the complex energy plane over a contour $C$
starting below the bottom of the valence band and ending at the Fermi energy,
and $g^{\sigma}_{ij}(z)$ are the site off-diagonal blocks of the system
Green function corresponding to a given geometry.
Possible lattice and/or layer relaxations at the overlayer are neglected.
The intersite Green functions $g^{\sigma}_{ij}(z)$ can be
evaluated either in the real space by using the cluster approach
\cite{Lie}, the recursion method \cite{Sonia,Haf}, or, as it is done
in the present paper and in Ref.~\onlinecite{Szu}, by the Bloch
transformation which employs the two-dimensional translational symmetry
of a given layer (for more details concerning the computational method see
Ref.~\onlinecite{book}).
We have calculated the EEI pairs $J_{ij}$ up to 101-shells of the fcc(001)
surface (i.e., up to the distance of about 10 $a$, where $a$ is the lattice
constant of the fcc lattice).
Such a large number of the EEIs is needed, in particular, for an accurate
estimate of the spin-wave stiffness constant in a real space, as it is also
known for the bulk case \cite{Antr,Marek}.
In actual calculations the sites $i,j$ were limited to the magnetic layer,
which is a good approximation in view of the smallness of the moments
induced in the Cu.
The spin-wave spectrum $E({\bf q}_{\|})$, the spin-wave stiffness constant
$D$, and the Curie temperatures $T_{c}^{MFA}$ and $T_{c}^{RPA}$ are
expressed, respectively, in terms of the EEIs as follows
\begin{eqnarray}
& & E({\bf q}_{\|})=\frac{4 \mu_{\rm B}} {M} \sum_{i \ne 0} \, J_{0i} \,
\left( 1-\exp(i{{\bf q}_{\|}} \cdot {{\bf R}_{i}}) \right) + \Delta \, ,
\nonumber \\
& & D=\frac{\mu_{\rm B}} {M} \sum_{ i \ne 0} \, J_{0i}R_{0i}^2 \, ,
\;\;\; k_{\rm B} T_{c}^{MFA}= \frac{2} {3} \, \sum_{i \ne 0} \,
J_{0i} \, ,
\nonumber \\
& & \frac{1}{k_{\rm B} T_{c}^{RPA}}= \frac{6 \mu_{\rm B}} {M} \,
\frac{1}{N_{\|}} \sum_{{\bf q}_{\|}} \, \frac{1} {E({\bf q}_{\|})} \, .
\label{eqs_EDT}
\end{eqnarray}
Here, $N_{\|}$ is the number of sites per layer,
the ${\bf q}_{\|}$-sum extends over the fcc(001) surface Brillouin
zone, $\mu_{B}$ is the Bohr magneton, $R_{0i}=|{\bf R_0 -R_i}|$ is
the interatomic distance, $M$ is the magnetic moment per atom, and
$\Delta$ is the magnetic anisotropy energy.
It should be noted that $T_{c}^{MFA}$ can be evaluated directly by
using the one-site rotation term $J_{0}$ (expressed in terms of the
site-diagonal element of the magnetic layer Green function similarly
as its bulk counterpart \cite{Lie,Marek}).
The expression for $T_{c}^{RPA}$ is a generalization of the bulk
counterpart \cite{RPA} to the case of magnetic layers: a vanishing
$T_{c}^{RPA}$ is obtained for $\Delta=0$ \cite{Bruno}
in an agreement with the Mermin-Wagner theorem \cite{MWT} and small
relativistic effects have to be considered in order to obtain a non-vanishing
value of $T_{c}^{RPA}$.
The anisotropy energy $\Delta$ is taken here as an adjustable parameter.
This is not a serious problem as the RPA Curie temperature has only a weak
logarithmic dependence upon $\Delta$ \cite{Bruno}, and it is thus sufficient
to know the order of magnitude of $\Delta$.
The latter is typically of the order of the dipolar energy $2 \pi M^{2}/V$,
where $V$ is the atomic volume.
In calculations we used $\Delta_{\rm Co}$=0.052~mRy  and
$\Delta_{\rm Fe}$=0.140~mRy.
 \begin{table}
 \caption{ Calculated values of effective exchange interactions for the
 first 10 shells of fcc(001) Fe- and Co-magnetic layers: a free-standing
 (fs) layer, an overlayer (ov) on fcc(001)-Cu, and an embedded (em) layer
 in fcc-Cu. Numbers of atoms in a given shell and the corresponding
 shell radii (in units of lattice constants) are given in parenthesis
 and square brackets, respectively.
 Corresponding values of magnetic moments $M$ are also given.}
 \begin{center}
 \begin{tabular}{rdddddd}
  Shell & \multicolumn{3}{c}{$J_s(Co)$ [mRy]} & \multicolumn{3}{c}
  {$J_s(Fe)$ [mRy]} \\
        &   fs   &   ov   &   em   &    fs  &    ov  &   em   \\ \hline
  1 (4) [0.71] &    2.85 &    2.34 &    2.01 &    3.40 &    2.69 &    2.62 \\
  2 (4) [1.00] &    0.24 &    0.14 & $-$0.12 &    0.12 &    0.22 &    0.39 \\
  3 (4) [1.41] & $-$0.02 & $-$0.06 & $-$0.01 & $-$0.39 & $-$0.37 & $-$0.30 \\
  4 (8) [1.58] & $-$0.03 &    0.05 &    0.03 & $-$0.23 & $-$0.13 & $-$0.05 \\
  5 (4) [2.00] &    0.03 &    0.01 & $-$0.01 &    0.05 &    0.03 &    0.04 \\
  6 (4) [2.12] & $-$0.01 & $-$0.07 & $-$0.07 & $-$0.01 &    0.15 &    0.20 \\
  7 (8) [2.24] & $-$0.02 &    0.00 &    0.00 &    0.09 & $-$0.03 & $-$0.07 \\
  8 (8) [2.55] &    0.00 &    0.00 &    0.00 &    0.01 &    0.04 &    0.03\\
  9 (4) [2.83] &    0.00 &    0.04 &    0.05 & $-$0.03 & $-$0.03 & $-$0.01 \\
 10 (8) [2.92] &    0.02 & $-$0.02 & $-$0.03 &    0.01 &    0.07 &    0.04 \\
 \hline
 $ M[\mu_B]      $ & 2.22 &  1.79  &  1.57  &  3.06  &  2.82  &  2.59  \\
 \end{tabular}
 \end{center}
 \label{Tab.1}
 \end{table}
The evaluation of $T_{c}^{RPA}$ is facilitated by observation that it is
proportional to the real part of the magnon Green function
$G_m(z)=N^{-1}_{\|} \sum_{{\bf q}_{\|}} \, (z-E({\bf q}_{\|})^{-1}$
corresponding to a dispersion law $E({\bf q}_{\|})$ and evaluated at $z=0$.
The corresponding ${\bf q}_{\|}$-summation is performed for complex energies
from which the value at $z=0$ is obtained by an analytic continuation
technique \cite{deconv}.
The sum for the evaluation of the spin-wave stiffness constant is
non-convergent due to the RKKY character of magnetic interactions
in metallic systems and to overcome this difficulty we have calculated
it by a regularization procedure \cite{etaf} described in detail in
Ref.~\onlinecite{Marek}.

The calculated EEIs for the first 10 shells, magnetic moments,
spin-wave stiffness constants, and the RPA and MFA Curie temperatures are
summarized in Tables~I and II for three limiting cases of magnetic Fe
and Co layers, namely the free-standing fcc(001) layer, the overlayer
on fcc-Cu(001) substrate, and the fcc(001) layer embedded in fcc-Cu host.

\begin{figure}
  \leavevmode
  \includegraphics{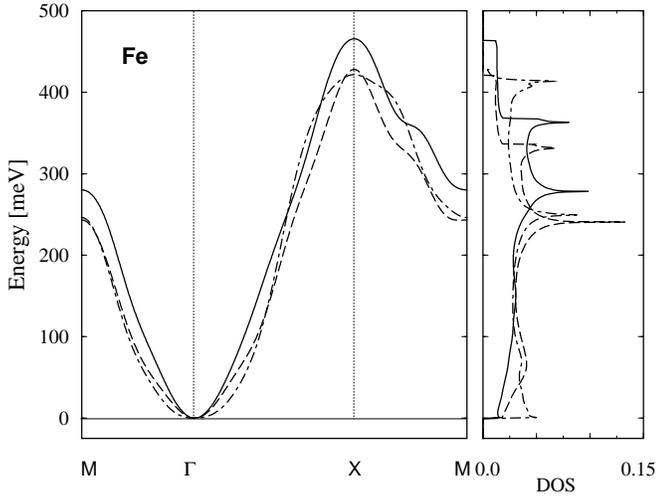}
 \vspace*{7.0cm}
 \caption{Magnon dispersion laws (left frame) and corresponding densities
 of states [in states/(meV$\cdot$\,atom] for the Fe-layer embedded in
 fcc-Cu (full line), Fe-overlayer on fcc-Cu(001) (dashed line), and the
 free-standing Fe-layer (dashed-dotted line). We have set here $\Delta = 0$.}
 \label{Fig.1}
 \end{figure}

Concerning the EEIs, the following general conclusions can be drawn:
(i) A pronounced dependence of magnetic moments on the coordination number
is found, namely their decrease with increasing number on nearest
neighbors, the effect being stronger for the Fe layer;
(ii) the EEIs are significantly enhanced (typically by a factor 2 or more)
as compared with their bulk counterparts;
(iii) the EEIs depend strongly upon the presence of a substrate and a
capping layer.
The latter dependence is due to the RKKY character of the EEIs in metals:
the coupling is not only mediated through the magnetic layer itself but
also via the substrate and capping layer.
This behavior is also clearly visible on the spin-wave spectra shown
in Fig.~1 for Fe and on the exchange stiffness constant (Table II).
We also present corresponding magnon densities of states (DOS) determined
from the magnon Green function $G_m(z)$.
A characteristic step of the height proportional to $1/D$ at the bottom of
the magnon DOS accompanied by a pronounced van Hove singularity in the
middle of the band are typical features of the two-dimensional bands
with the nearest-neighbor interactions which are here only slightly
modified by non-vanishing interactions in next shells.
Interestingly,
the spin-wave stiffness constants and Curie temperatures behave differently
as a function of the atomic coordination for Co and Fe layers, i.e.,
for cases of the free-standing layer, the overlayer, and the embedded
layer.
This behavior can be related to the values of leading EEIs in both cases,
in particular to large antiferromagnetic couplings of 3rd and 4th
nearest neighbors of Fe-based layers which effectively reduce the
value of the spin-wave stiffness constant (see Eq.~(\ref{eqs_EDT})), in
particular for the free-standing layer.
On the contrary, the Co-based EEIs have the prevailing ferromagnetic
character giving thus increasing spin-wave stiffness constants due to
increasing values of the EEIs with reduced atomic coordination.
The antiferromagnetic character of the EEIs for fcc-based Fe layers is
strongly enhanced as compared to the bcc-Fe case \cite{Marek} while
the prevailing ferromagnetic character of the EEIs for bulk fcc-Co
\cite{Marek} and for fcc-Co layers remains unchanged.

\begin{figure}
  \leavevmode
  \includegraphics{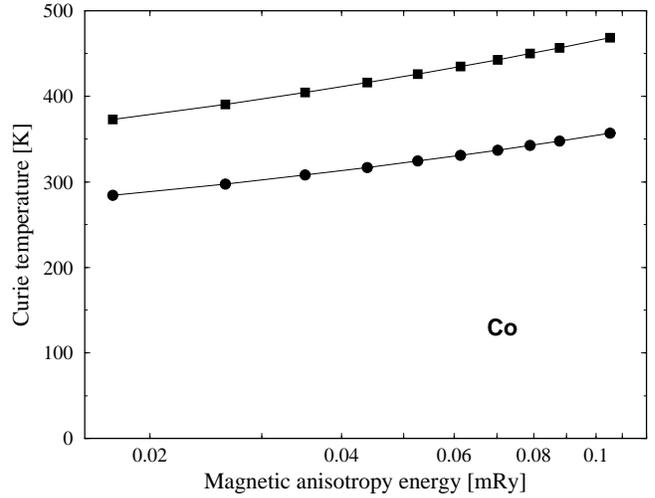}
  \vspace*{7.0cm}
\caption{ RPA Curie temperatures of Co overlayer on fcc-Cu(001) (full squares)
and of fcc(001)-Co layer embedded in fcc-Cu (full circles) as a function of
the magnetic anisotropy energy $\Delta$. Note the logarithmic scale on the
abscise.}
\label{Fig.2}
\end{figure}

The MFA Curie temperatures are typically of the same order magnitude as
the corresponding bulk ones due to the fact that the reduced coordination
is approximately compensated by the increase of the EEIs.
This observation is in a strong disagreement with experimental data:
this failure is due to the fact that the MFA violates the Mermin-Wagner
theorem due to the neglect of collective transverse fluctuations
(spin-waves) and it is thus inappropriate for two-dimensional systems.

\begin{figure}
  \leavevmode
  \includegraphics{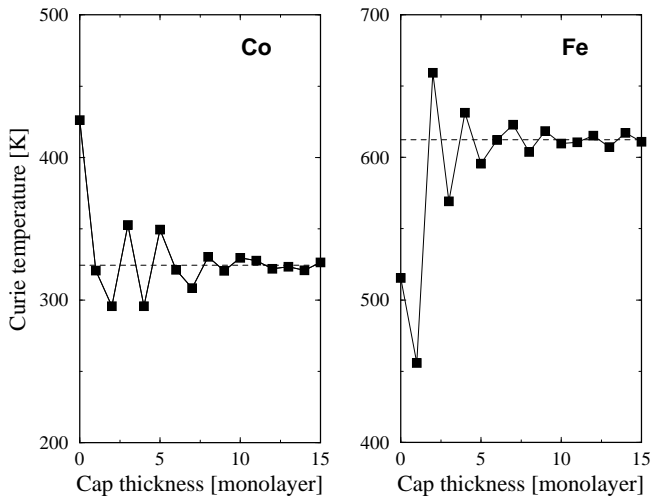}
  \vspace*{7.0cm}
\caption{ $T_{c}^{RPA}$ of a Co (left) and Fe (right) overlayer on
fcc-Cu(001) substrate covered by a cap layer of varying thickness.
The dashed lines represent the embedded layer limit (infinite cap
thickness) while the limit of zero cap thickness corresponds to the
uncovered overlayer.}
\label{Fig.3}
\end{figure}

\begin{table}
\caption{ Calculated values of the spin-wave stiffness coefficient $D$,
$T_{c}^{RPA}$, and $T_{c}^{MFA}$ for Fe and Co magnetic layers:
a free-standing (fs) layer, an overlayer (ov) on fcc(001)-Cu, and
an embedded (em) layer in fcc-Cu.}
\begin{center}
\begin{tabular}{lllllll}
 & \multicolumn{2}{c}{ $D$ [meV$\cdot$\AA${}^2$]}
& \multicolumn{2}{c}{ $T_C^{RPA}$ [K]}
& \multicolumn{2}{c}{ $T_C^{MFA}$ [K]}  \\
   & Fe & Co & Fe & Co & Fe & Co \\ \hline
fs & $164\pm4$  & $570\pm13$ & $400$ & $529$ & $1265$ & $1300$ \\
ov & $331\pm14$ & $532\pm9$  & $515$ & $426$ & $1068$ & $1043$ \\
em & $462\pm16$ & $416\pm8$  & $612$ & $324$ & $1189$ & $\ \, 797$  \\
\end{tabular}
\end{center}
 \label{Tab.2}
\end{table}

The RPA Curie temperatures as a function of the anisotropy energy $\Delta$
are shown in Fig.~2 for cases of Co overlayer on fcc-Cu (001) and fcc-Co(001)
layer embedded in Cu.
The weak logarithmic dependence of $T_c^{RPA}$ on $\Delta$ \cite{Bruno}
is obvious: $T_c^{RPA}$ varies by about 25\% as $\Delta$ varies by
an order of magnitude so that the results are not significantly
influenced by our semi-empirical choice of $\Delta$.
The RPA Curie temperatures are strongly reduced as compared to the
corresponding bulk values thereby improving on the MFA results.
Nevertheless, they are still too large as compared to observed Curie
temperatures of ferromagnetic monolayers (being of order 150~$-$~200 K).
It is unclear whether this is due to some inaccuracy of the theory or to
some imperfections of the samples used in experiments.
On the contrary, such important experimental facts as the strong influence
of the metallic coverage on the Curie temperature \cite{Kirsch} are well
explained by our theory as illustrated in Fig.~3.
The oscillatory character of $T_{c}^{RPA}$ around the value corresponding
to an infinite cap, i.e., to the limit of the embedded layer, is clearly
visible and it is in a qualitative agreement with the recent experiment of
Vollmer {\em et al.\/} \cite{Kirsch}.
The origin of these oscillations can be traced back to the oscillatory
behavior of the EEIs and it has the same origin as related oscillations
of the interlayer exchange couplings found for the Co/Cu/Co(001)
trilayer with a varying Cu cap-layer thickness \cite{OI}.
These oscillations are due to quantum-well states in the Cu-cap layer
formed between the vacuum and the magnetic layer which, in turn,
influence properties of the magnetic layer.
We have verified that amplitudes of oscillations of the EEIs decay with
the thickness $d$ of the cap layer approximately as $d^{-2}$.
The same thickness dependence was also found for the related case of
the interlayer exchange interactions for the Co/Cu/Co trilayer with
the varying thickness of the Cu-cap layer \cite{IECcap}.
A similar behavior was also verified for the oscillatory dependences
of $T_{c}^{RPA}$ and $T_{c}^{MFA}$ which, in turn, are derived from
the EEIs.
It should be noted that amplitudes and phases of oscillations can
be influenced by the thickness of magnetic layer and/or the presence
of the disorder in the system.

In conclusion, in view of the interpretation proposed here,
the oscillatory behavior of the Curie temperature of Fe
films as a function of the Cu-cap thickness as reported by Vollmer
{\em et al.\/} \cite{Kirsch} would constitute
the first {\em direct\/} experimental evidence of the
oscillatory RKKY character of exchange interactions in itinerant
ferromagnets.

J.~K., I.~T., and V.~D. acknowledge the financial support provided
by the Grant Agency of the Czech Republic (No.\ 202/00/0122),
the Academy of Sciences of the Czech Republic (No.\ A1010829),
and the Czech Ministry of Education, Youth and Sports
(OC P3.40 and OC P3.70).


\begin{references}
\bibitem{etaf} The procedure consists in introducing an exponential damping
factor $\exp(-\eta R_{0i}/a)$ which makes the sum absolutely convergent and
subsequently taking the limit $\eta \rightarrow 0$. The error bars for $D$
in Table II correspond to the uncertainty due to this latter step.
\end{references}
\end{document}